\RequirePackage{fix-cm}
\documentclass[smallextended]{svjour3}       
\smartqed  
\usepackage{graphicx}
\usepackage{amsmath}
\usepackage[strings]{underscore}  

\begin{document}

\title{New developer metrics:  
}
\subtitle{Are comments as crucial as code contributions?}

\titlerunning{New developer metrics}        

\author{Abdulkadir \c{S}eker \and Banu Diri \and Halil Arslan
}

\authorrunning{A. \c{S}eker, B. Diri, et al.} 

\institute{Abdulkadir \c{S}eker \at
	Sivas Cumhuriyet University \\
	\email{aseker@cumhuriyet.edu.tr}      
	\and
	Banu Diri \at
	Y{\i}ld{\i}z Technical University
	\email{diri@yildiz.edu.tr}
	\and
	Halil Arslan \at
	Sivas Cumhuriyet University \\
	\email{harslan@cumhuriyet.edu.tr}   
}

\date{Received: date / Accepted: date}

\maketitle

\begin{abstract}
Open-source code development has become widespread in recent years. As a result, open-source software platforms have also become popular, and millions of developers from diverse locations are able to contribute to the same projects. On these platforms, various knowledge about them is obtained from user activity. This information is used in the form of developer metrics to solve a variety of challenges. In this study, we proposed new developer metrics, including commenting and issue-related activity, that require less information. We concluded that commenting on any feature of a project can be equally as valuable as code contribution. In addition, besides the quantitative ones, metrics based on only the existence of the activity have been shown to offer also considerable results. We saw that issues were crucial in identifying user contributions. Even if a developer makes a contribution to only one issue on a project, the relation between the developer and the project is tight. The hit scores are relatively lower because of the sparsity problem of our dataset; even so, we believe that we have presented improvable and remarkable new developer metrics.

\end{abstract}

\section{Introduction}
\label{intro}
Thanks to the increasing capabilities of open source software development tools, the number of open source users and projects is growing each year. These platforms include millions of developers, each of whom has a different character and skill set, as well as a wide variety of projects that offer solutions to different problems. In environments with such a large amount of data, it is also difficult for developers to find similar products to their own, discern projects of interest, and reach projects to which they can contribute. As developers primarily use search engines or in-platform search menus to find projects, the constraints of text-based search \cite{McMillan2012} and challenges related to finding the correct keywords also cause some projects to be missed \cite{Hu2015}.  While various project recommendation systems are being developed to overcome this problem, projects must be rated by users for recommendation models to work properly. In the same way that viewers give ratings to movies that they have watched, developers need to rate the projects in which they are interested. However, this is not currently the case on (open-source) software development platforms. Several software and developer metrics are used to calculate the score that a user gives projects, which are extracted from the activity or features of both developers and projects.

Developer metrics which are used in many challenges include the number of lines of code, developers’ degrees of connection to one another, past experience, or common features (nationality, location, occupation, gender, previously used programming languages, etc.). These metrics offer solutions to different problems within open source software development and distributed coding, including automatic assignments (task, issue, bug, or reviewer) \cite{DeLima2015,Badashian2015,Junior2018}, project recommendation systems \cite{Zhang2014,Sun2018}, software defect detection \cite{OzcanKini2018}, etc.

In this study, new developer metrics are presented to be used for different problems. We developed a project recommendation system for the evaluation of metrics and obtained remarkable results. The recommendation system was developed based on a dataset consisting of data from GitHub. Most GitHub users are familiar with (i.e., contribute to or interest) relatively few projects hosted on the platform. Because of this, a critical sparsity problem has occurred. To address the problem, we selected a dataset with a high users-projects ratio.

The structure of this paper is as follows. In the background section, we describe the literature on previously proposed metrics and project recommendation models for open-source software development platforms. In the research design section, the dataset used in this study is introduced and the proposed metrics are detailed. In the final section, the proposed metrics are assessed in terms of their accuracy rates in the project recommendation system.

\section{Background}
\label{sec:1}

Pull request (PR) allows users to inform others about changes they have pushed to a branch in a repository on GitHub. PRs are a key feature in contributing code by different developers to a single project \cite{Gousios2014}. The proposed metrics related to this feature are used to solve different PR problems. PRs need to be reviewed by a reviewer to merge projects. If the result of a revision is positive, the PR is integrated into the master branch. Finding the correct reviewer is an important parameter for ensuring rapid and fair PR revisions. In this context, different metrics have been used to address the problem of automatic PR reviewer assignment. Existing literature has proposed various metrics to solve this problem, such as PR acceptance rate within a project, active developers on a project  \cite{Junior2018}, PR file location  \cite{Thongtanunam2015}, pull requesters’ social attributes \cite{Tsay2014}, and textual features of the PR  \cite{Yu2014}, among others. 

Closing a PR with an issue, PR age, and mentioning (@) a user in the PR comments have all been used to determine the priority of a PR \cite{VanDerVeen2015}. Cosentino proposed three developer metrics (community composition, acceptance rates and become a collaborator) to investigate project openness and stated that project owners could evaluate the attractiveness of their projects using these metrics \cite{Cosentino2014}.  

Developer metrics are also used in the detection of software defects. In one study, defects were estimated using different metrics grouped by file and commit level. The number of files belonging to a commit, the most modified file of all files in a commit, the time between the first and last commits to a file, and the experience of a given developer on a committed file were identified as important metrics \cite{OzcanKini2018}.

Reliability metrics are used to quantitatively express the reliability of a software product \cite{Kaur2014}. To measure reliability in open source projects, metrics such as number of contributors, number of commits, number of rows in commits, and certain metrics derived from them are used. Tiwari et al. proposed two important metrics for reliability: contributors and number of commits per 1000 code lines \cite{Tiwari2012}.

Code ownership metrics are also important for open source software. One study used the modified (touched) number of files to rank developer ownership according to code contributions  \cite{Bird2011}. In another study, the number of changed lines in a file (churn) was used to address this problem \cite{Munson1998}. Foucault confirmed the relationship among these code ownership metrics and software quality \cite{Foucault2015}.

Recommender systems are an important research topic in software engineering \cite{Happel2008,Robillard2010}. In ordinary recommendation models, previously known user-item matrices are used; in other words, the rating given by a user for an item is known. In this state, the essential research topic is on estimating with different algorithms and models the rating that the user has already given \cite{Sharma2013}. However, the point in question is different on open source software platforms. Considering the developer as the user and the project (repository) as the item, the rating given by a developer to a project is unknown. In this context, the first problem that must be solved is how to create an accurate developer-project matrix. At this point, different developer metrics come into play.

GitHub is the world’s largest code server, hosting more than 40 million repositories to which over 100 million developers have contributed\footnote{https://en.wikipedia.org/wiki/GitHub}. As such, GitHub is a reasonable choice for developing a project recommendation model. Sun et al. relied on basic user activity to develop a model using GitHub data. Specifically, when rating a project for a developer, they used like-star-create activities related to projects \cite{Sun2018}. We also used this scoring method to compare with our metrics in this paper. In another study, a developer’s social connections, programming with a common language, and contributions to the same projects or files were used as metrics \cite{Casalnuovo2015}. Liu designed a neural network–based recommendation system which used metrics such as working at the same company, previous collaboration with the project owner, and different time-related features of a project \cite{Liu2018}. In a study aiming to predict whether a user would join a project in the future, the metrics used included a developer’s GitHub age (i.e., when their account was opened), the number of projects that they had joined, the programming languages of their commits, how many times a project was starred, the number of developers that joined a project, and the number of commits to a project \cite{Nielek2017}.

\section{Research Design}
\label{sec:2}

\subsection{Dataset}
One of the most serious challenges in developing a recommender system is sparsity \cite{Niu2016}, a problem that occurs when most users rate only a few items \cite{Guo2012}. This issue is present on GitHub because it is not possible for developers to be aware of the millions of repositories on the platform. In the studies mentioned in the previous section, we observed that limited (less sparse) data were used, which is contrary to the nature of the GitHub environment. Although the results obtained in these studies appear successful, the question remains how successful the proposed algorithms will be on real platform data. In light of this, a sub-dataset reflective of the sparsity problem inherent to GitHub was used in this study. The dataset contained all data related to 100 developers and 41,280 projects \cite{Seker2020}. The creators of the dataset indicated that they selected the most active users on the platform. They then extracted all related data for these users from GitHub (commits, issues, pull requests, comments about these activities, watchers, etc.). Thus, we anticipated that the recommender system we developed would produce results parallel to those for the larger dataset of the platform as a whole. In this regard, although our evaluation results seem weak compared to similar studies, we believe that the proposed metrics are worthy of consideration.

 \subsection{Project Recommendation System}
\begin{figure*}[ht]
	\includegraphics[width=0.75\textwidth]{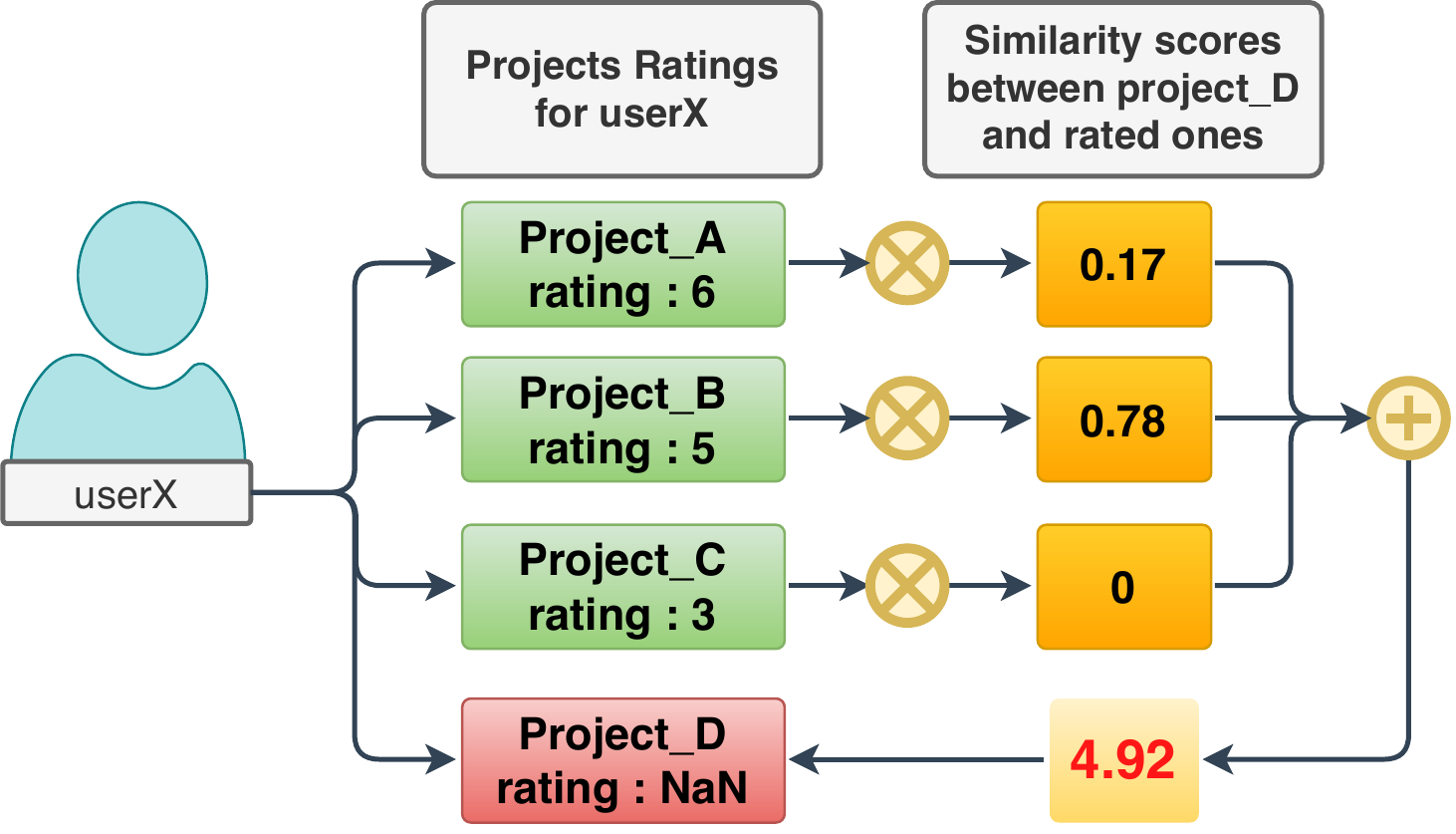}
	\caption{Calculating unrated project with the help of similarity rated projects.}
	\label{fig:similarity}       
\end{figure*}

Designing a project recommender system for open source software development platforms includes two stages. First, the project-developer rating matrix is generated using specific metrics. Second, the top-k projects are recommended to each developer. Finally, the accuracy of the suggestions is evaluated. In this study, the recommendation model was designed as follows:

\begin{enumerate}
\item Different developer metrics were used to obtain the score matrix. The values of the features (metrics) were scaled from 0 to 10. As in the movie-user model, each developer will thus have given a rating (0–10) to each project.
\item The similarity between unknown projects\footnote{We assumed that “unknown” projects were those to which developers had no relationship and had made no contributions} and rated projects was used to calculate the rating of unknown projects \cite{Sun2018}. The similarity value between the projects was calculated using cosine similarity. When calculating the rating of an unknown one, the dot product of the similarity values between the projects that the user rated and the unknown project was used (Equation \ref{equ:score}). An example scenario involving this calculation is presented in Figure \ref{fig:similarity}.

\begin{equation}
\label{equ:score}
	unknown_{rating} = \sum_{n=0}^{i} known_{rating}^i * similarity_{known^i,unknown} 
\end{equation}

\item The top 5 highest-rated projects among the unknown projects were recommended to each developer.
\item The accuracy of the recommendations was evaluated.
\end{enumerate}

\subsection{Evaluation Techniques}

While recommending projects to developers, there should be a ground truth for evaluating the proposed projects. Unlike ordinary recommender systems, there is an unsupervised model. The evaluation criteria for some studies related to this subject are set forth below.
 
\begin{itemize}
\item A user-project-rating matrix was split randomly into test and training subsets. Accuracy or recall scores were then calculated from the intersection between the top-n scores of the test and training subsets \cite{Sun2018}. However, another study stated that this method should not be used on platforms like GitHub where time is an important parameter, pointing out that the problem of predicting past activity with future data will occur when using k-fold cross-validation by dividing the data randomly \cite{Junior2018}.
\item In another study, the recommended projects accuracy was evaluated using the developer’s past commits to the related project. A recommendation was assumed to be correct if the number of commits belonging to a certain developer on the project was over a certain value. The average number of commits per project was set as the threshold value in the dataset \cite{Liu2018}.
\item In a study predicting whether a developer would join a project in the future, the dataset was split into two different sets by time. In this way, the predicted result was verified with actual future data \cite{Nielek2017}.
\end{itemize}

In this study, GitHub’s “watching” feature was used as the ground truth. GitHub users can follow, or “watch,” projects whose developments they want to monitor \cite{Sheoran2014}. If a developer is watching a project, this indicates that he/she is interested in the project. Thus, “watching” can be considered as a real evaluation criterion. In our model, the top-n projects were recommended to each developer. If the recommended projects were among the developer’s watched projects, the projects were considered a hit\footnote{In other words, it is a correct recommendation}. The case of a developer watching fewer than n projects was taken into account in the score equation \ref{equ:hitscore}.

\begin{equation}
\label{equ:hitscore}
Hit_{score} = 
\begin{cases}
100*\frac{hit_{fullname} + (hit_{owner}*0.5)}{n}, & \text{if } num_{watched}\geq n\\ \\
100*\frac{hit_{fullname} + (hit_{owner}*0.5)}{num_{watched}},              & \text{otherwise}
\end{cases}
\end{equation}

The full name of a GitHub repository (project) is created by concatenating the owner’s username with the repository name. In analyzing our results, we noticed that the model recommended a project to a developer that only hit the owner’s name—that is, the model found an incorrect project by the correct owner. We evaluated this proposal has half the correct score, as recommending the correct owner to a developer will allow the developer to access the owner’s other projects.

An example scenario demonstrating this situation is given in Table \ref{tab:recomOwner}. The projects recommended for Alice are listed in the first column. Since four of them are among the projects that Alice watches, the initial score is \textit{4}. In addition, there are two projects by a developer named "\textit{fengmk2}" among Alice’s watched projects (\textit{“fengmk2/parameter”} and \textit{“fengmk2/cnpmjs.org”}). For the fourth proposed project \textit{“fengmk2/emoji”}, the owner’s name was guessed correctly, but the repository name was missed. In this case, Alice will be aware of other projects by  \textit{"fengmk2"}.  Thus a half-point is added to the initial score and \textit{4.5} is the final score. 

\begin{table}[ht]
	\caption{A sample that is recommended correctly of the only the project owner }
	\label{tab:recomOwner}       
\begin{tabular}{lll}
	\hline\noalign{\smallskip}
	\textbf{Top-5 Recommendation}                     & \textbf{Full\_name   Matchs}      & \textbf{Owner matchs} \\ 
	\hline\noalign{\smallskip}
	 iojs/io.js                      &iojs/io.js                            &                                         \\
    juliangruber/co-read  & juliangruber/co-read     &                                          \\
    koajs/compose          &koajs/compose               &                                           \\
    \textbf{fengmk2/emoji} &                                    &\textbf{fengmk2}             \\
    visionmedia/co         &visionmedia/co                &                                         \\ 
    \hline\noalign{\smallskip}
\end{tabular}
\end{table}

\begin{figure*}[ht]
	\includegraphics[width=0.85\textwidth]{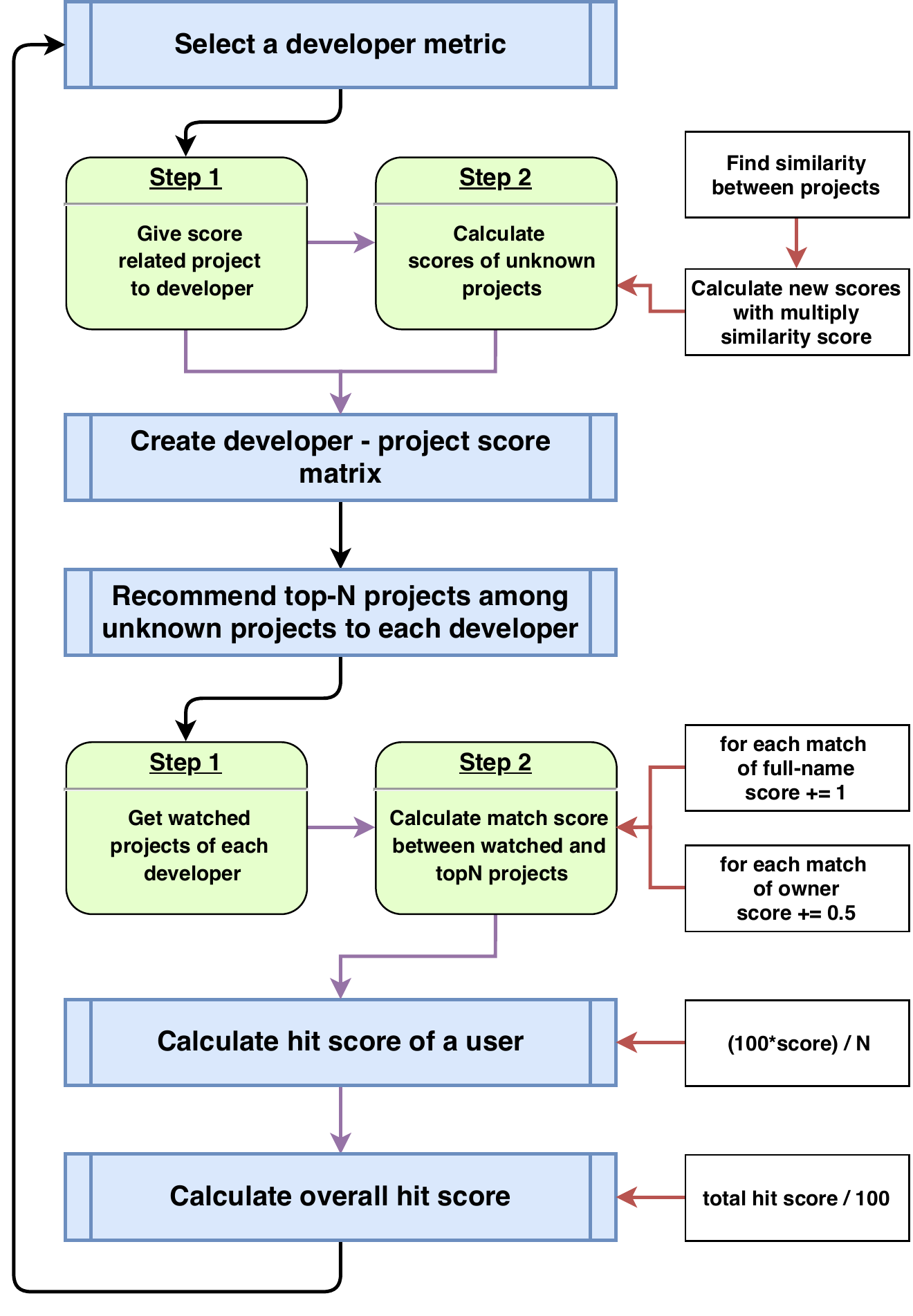}
	\caption{Project recommending model flowchart}
	\label{fig:algorithm}       
\end{figure*}

In this way, a project recommendation model has been created for open source platforms. The algorithm of the recommendation model is presented in Figure \ref{fig:algorithm}, starting with selecting a feature as a metric and end with calculating hit scores.

\section{Empirical Results}
In this section, different developer metrics are given. These metrics provide information about a developer’s past activity on a project. All metrics used were scaled from 0 to 10 using the min-max normalization technique. The developer-project relationship was thus rated in the range 0–10 (as with a viewer’s rating of a movie). The results were calculated for the top 5 recommendation hit scores. We experimented with several metrics, including those based on coding language; obtained from the ratio of how many times a developer performed an activity to the total number of the activity; and created using different normalization methods. However, only metrics that obtained hit scores greater than 5\% are shown in this study (except single metrics). 

\subsection{Single Metrics}
Developer activity on projects was handled as a metric. Activity includes all kinds of comments, code contributions, revisions, and so on. In this section, all metrics were used individually in order to evaluate the significance of each. These metrics refer to the number of activities per project for a given developer (Table \ref{tab:singlemetrics}).
\begin{table}[ht]
	\caption{Single developer metrics}
	\label{tab:singlemetrics}       
	\begin{tabular}{lll}
		\hline\noalign{\smallskip}
		& Metric                 & Definition                                         \\
		\hline\noalign{\smallskip}
		1  & issue\_opened              & Number of issue opened                   \\
		2  & issue\_commented      & Number of comments to issues       \\
		3  & issue\_closed               & Number of issue closed                    \\
		4  & issue\_closedwithPR   & Number of issues closed with a PR  \\
		5  & issue\_assigned           & Number of issues assigned              \\
		6  & commit\_commented  & Number of comments to commits    \\
		7  & commit\_authored       & Number of authorship in commits    \\
		8  & commit\_committed    & Number of commits                             \\
		9  & pr\_opened                   & Number of PR opened                        \\
		10 & pr\_merged                  & Number of PR merged                         \\
		11 & pr\_assigned                & Number of PR assigned                        \\
		12 & pr\_commented          & Number of comments to PRs               \\
		
		\hline\noalign{\smallskip}     
	\end{tabular}
\end{table}

All single metrics were given to the model individually, and the scores presented in Table \ref {tab:singlemetricscore} were obtained according to the evaluation technique outlined above. In addition to our metrics, another metric that extracted from the study of Sun et al. was added to make a comparison. They scored developers and projects using like-star-create activities. They used text data extracted from projects’ README and source code files to find project similarities \cite{Sun2018}. There were approximately 22,000 repositories and 1,700 developers in their dataset, which was created using data from four groups of projects. 

We planned to use this less sparsed dataset to make a fair comparison but could not because the dataset was unshared. Owing to we could not communicate with them, we applied their rating algorithms to our dataset.

\begin{table}[ht]
	\caption{Single developer metrics scores} 
	\label{tab:singlemetricscore}       
	\begin{tabular}{llll}
			\hline\noalign{\smallskip}
		Metric                   			& Hit Score (\%)  \\
		\hline\noalign{\smallskip}
		issue\_commented  		& 15.3         	      \\

		issue\_closedwithPR 	& 15        	        \\
		issue\_opened       		& 14        	       \\
		pr\_opened          			& 13.7   	         \\
		commit\_commented   & 11.9      		      \\
		pr\_commented       		& 10.9    	         \\
		pr\_merged          			& 9.3       	        \\
		\textbf{Sun's metric}		& \textbf{7.7}   \\ 
		commit\_authored   		 & 6           			\\
		commit\_committed   	& 5.7       	      \\
		issue\_closed       			& 3        	 	      \\
		issue\_assigned    			 & 2.8     		      \\
		pr\_assigned        			& 2.5      		      \\
		\hline\noalign{\smallskip}
	\end{tabular}
\end{table}

When the results are analyzed, it is clear that the issue-related metrics are crucial even by themselves. Closing an issue with a PR means that the PR fixed a bug or issue in the project \cite{VanDerVeen2015}. As our results indicate, \textit{issue\_closedwithPR} is a remarkable metric. Opening an issue or PR is also an important metric. The most interesting conclusion that can be drawn from these results is that comments have higher hit scores than direct code contributions.

\subsection{Fusion Metrics}
In these results, we observed that some metric groups came to the forefront. New metrics can be proposed by grouping comments, code contributions, or other common featured metrics. In this context, fusion metrics were created from single metrics.

\begin{enumerate}
	\item\textit{count:} is created from the sum of all metrics.
	 
	\item \textit{contribution:}  is created from the sum of all code contribution-related metrics.
	\[contribution = pr\_opened + issue\_opened + issue\_closedwithPR + pr\_merged + commit\_committed \]
	
	\item \textit{comment:} is created from the sum of all comment-related metrics.
	\[comment =	issue\_commented + commit\_commented+ pr\_commented \]
	
	\item \textit{issue\_related:}  is created from the sum of all issue-related metrics. 
	\[issue\_related =  issue\_opened + issue\_closedwithPR + issue\_commented + issue\_assigned  \]
	
	\item \textit{pr\_related}  is created from the sum of all PR-related metrics.
	\[pr\_related = pr\_opened + pr\_merged + pr\_closed +  pr\_assigned \]
	
	\item \textit{commit\_related:}  is created from the sum of all commit-related metrics.
	\[commit\_related =	commit\_commented + commit\_authored + commit\_committed \]
	
	\item \textit{commit2comment}  is created from the (\textit{commit\_committed} divided by \textit{commit\_commented})
	\[commit2comment =	\frac{commit\_committed}{commit\_commented} \]
	
	\item \textit{issue2comment}  is created from the (\textit{issue\_opened} divided by \textit{issue\_commented})
	\[issue2comment =	\frac{issue\_opened}{issue\_commented} \]

	\item \textit{pr2comment}  is created from the (\textit{pr\_opened} divided by \textit{pr\_commented})
    \[pr2comment =	\frac{pr\_opened}{pr\_commented} \]
	
	\item \textit{code2comment}  is created from the ratio of two fusion metrics (\textit{contribution} divided by \textit{comment})
	\[code2comment =	\frac{contribution}{comment} \]

\end{enumerate}

The results of the fusion metrics are presented in Table  \ref{tab:fusionmetricscores}.a. Most fusion metrics had a positive impact on hit score. \textit{comment} in particular was a remarkable metric which showed that commenting is as important as code contributions. In addition, the \textit{issue\_related} also drew our attention. The results of the ratio-based metrics were not as successful as the others. 

\begin{table}[ht]
	\caption{Fusion developer metrics scores} 
	\label{tab:fusionmetricscores}       
	\begin{tabular}{ll}
		\multicolumn{2}{c}{(a)}  \\
		\hline\noalign{\smallskip}
		Fusion Metric                   	& Hit Score (\%)  \\
		\hline\noalign{\smallskip}
		comment        					& 17        \\
		issue\_related  					& 16          \\
		contribution   					& 15.6       \\
		count          						    & 15.2       \\
		pr\_related    						 & 14.9       \\
		issue2comment					& 14.2    \\
		code2comment					& 13         \\
		commit\_related 				& 11       \\
		pr2comment 				& 11       \\
		\textbf{Sun's metric	}	& \textbf{7.7}   \\ 
		commit2comment				& 6.9       \\
		\hline\noalign{\smallskip} 
	\end{tabular}
\&
	\begin{tabular}{ll}
		\multicolumn{2}{c}{(b)}  \\
		\hline\noalign{\smallskip}
		Binary Fusion Metric                   	& Hit Score (\%)  \\
		\hline\noalign{\smallskip}
		issue\_related        					& 20        \\
		count  										& 19         \\
		contribution   					  & 18.8       \\
		comment          					 & 18.4       \\
		pr\_related    						 & 15.5       \\
		commit\_related 				& 11.6       \\
		\textbf{Sun's metric	}	& \textbf{7.7}   \\ 
		issue2comment					& x    \\
		pr2comment						& x    \\
		commit2comment 				& x       \\
		code2comment					& x         \\
		\hline\noalign{\smallskip} 
	\end{tabular}
\end{table}	

\subsection{Binary Metrics}
We tried different metrics for the project recommendation in the previous section. It is interesting that the metric consisting only of comments achieved a higher success rate than the metric created by collecting all single metrics. We continued to study different metrics to analyze whether success could be increased with less information. 

In the above single metrics, there is information about how many activities were made (For instance, John opened 18 issues in the projectX, the rating of John-projectX is 18). Alternatively, a set of metrics was created simply showing whether that activity existed (For instance, even so, John opened an issue on the projectX, the rating of John-projectX is 1 or there is no opened issue by John in the projectY, the rating of John-projectY is 0). In this context, the \textit{binary metrics} were created using the equation \ref{equ:binarymetric}.

\begin{equation}
 \label{equ:binarymetric}
	Binary Metrics= 
	\begin{cases}
	1,& \text{if } Single Metric > 0 \\
	0,              & \text{otherwise}
	\end{cases}
\end{equation}

\subsubsection{Binary Fusion Metrics} 
Binary metrics and ratio-based fusion metrics consist only of 0s and 1s. As such, using them directly will not generate logical results. Therefore, binary fusion metrics were generated from these metrics, and only the results of these binary fusion metrics are given (Table \ref{tab:fusionmetricscores}.b).

The results improved, and the top success metrics rankings swapped places. The best five metrics were the same as in the previous section. These results show that the number of comments is crucial in using comments as a metric. So, the presence of comments alone is not sufficient to use it properly. On the other hand, the issue\_related metric is leading among the binary metrics, indicating that issues are the most important feature for the developer-project relationship. Apart from our findings regarding the importance of comments, obtaining better results from issues than commits was one of the most surprising results of this study.

\section{Conclusion}
The main purpose of this study was to propose new developer metrics that could be used to solve different software engineering challenges. In this study, the features extracted from user activity on an open source platform (GitHub) were used. The study focused on finding metrics that would enable greater success with less knowledge. Some fusion metrics gave successfull results.

It is clearly seen that the \textit{comment}  made gave impressive results. For this metric, quantity is an important parameter. It means the more a developer writes comments, the more related to the project. On the contrary, it can deduce from the results of binary fusion metrics that the\textit{ issue\_related } is a quantity-free metric. It also means, to use this metric is adequate to know it's presence is whether or not. On this regard, It is revealing that \textit{issue} is a significant feature for open-source platforms. In addition, metrics that used features based on existence (binary metrics) were highly successful, showing that for some activities, there is no need for quantities in order to extract knowledge.

We presented these new developer metrics, but we are curious why some of them became prominent. In light of this, we are planning another study involving a survey for junior and senior developers whom we can contact to understand the ground truth of our metrics’ success (especially the \textit{comment} metric).

Because of the sparsity problem, our hit scores may not higher enough when comparing the other similar studies. Even so, we think that we offered some new improvable developer metrics. Moreover, to compare our results, we added a very similar metric used in Sun et al.’s study \cite{Sun2018}, which revealed that most of our metrics surpassed that metric. In this context, we plan to apply the obtained metrics to different datasets for validity. 

In addition, we anticipate that these metrics will be useful for solving various problems. Many developers besides owners and collaborators have contributed to projects due to the open source nature of GitHub. On some projects, external developers even made more contributions than the core team. These metrics can reveal developers’ contribution rankings on a project. To implement this, we plan to cooperate with a software company. 

%
%

%


%
%

\bibliographystyle{spmpsci}      
\bibliography{metrics.bib}   

\end{document}